\documentclass[sigconf]{acmart}

\usepackage{caption}
\usepackage{subcaption}
\usepackage{url}
\usepackage{amsmath}
\usepackage{multirow}
\usepackage{makecell}
\usepackage[flushleft]{threeparttable}
\usepackage{verbatim}
\usepackage{float}
\usepackage{bm}

\AtBeginDocument{%
  \providecommand\BibTeX{{%
    \normalfont B\kern-0.5em{\scshape i\kern-0.25em b}\kern-0.8em\TeX}}}

\newcommand{\shortname}{RGCL}
\newcommand{\fullname}{Review-aware Graph Contrastive Learning~}

\newcommand{\fusingshortname}{RG}
\newcommand{\fusingfullname}{Review-aware Graph learning~}

\copyrightyear{2022} 
\acmYear{2022} 
\setcopyright{acmcopyright}
\acmConference[SIGIR '22]{Proceedings of the 45th International ACM SIGIR Conference on Research and Development in Information Retrieval}{July 11--15, 2022}{Madrid, Spain.}
\acmBooktitle{Proceedings of the 45th International ACM SIGIR Conference on Research and Development in Information Retrieval (SIGIR '22), July 11--15, 2022, Madrid, Spain}
\acmPrice{15.00}
\acmISBN{978-1-4503-8732-3/22/07} 
\acmDOI{10.1145/3477495.3531927}

\begin{document}
\fancyhead{}

\title{A Review-aware Graph Contrastive Learning Framework for Recommendation}

\author{Jie Shuai}
\affiliation{
\department[0]{Key Laboratory of Knowledge Engineering with Big Data,}
\institution{Hefei University of Technology}
\country{}
}
\email{shuaijie.hfut@gmail.com}

\author{Kun Zhang}
\affiliation{
\department[0]{Key Laboratory of Knowledge Engineering with Big Data,}
\institution{Hefei University of Technology}
\country{}
}
\email{zhang1028kun@gmail.com}

\author{Le Wu}
\authornotemark[1]
\affiliation{
\department[0]{Key Laboratory of Knowledge Engineering with Big Data,}
\institution{Hefei University of Technology}
\institution{Institute of Artificial Intelligence, Hefei Comprehensive National Science Center}
\country{}
}
\email{lewu.ustc@gmail.com}
\thanks{Le Wu is the Corresponding author.}

\author{Peijie Sun}
\affiliation{
\department[0]{Key Laboratory of Knowledge Engineering with Big Data,}
\institution{Hefei University of Technology}
\country{}
}
\email{sun.hfut@gmail.com}

\author{Richang Hong}

\affiliation{
\department[0]{Key Laboratory of Knowledge Engineering with Big Data,}
\institution{Hefei University of Technology}
\institution{Institute of Artificial Intelligence, Hefei Comprehensive National Science Center}
\country{}
}
\email{hongrc.hfut@gmail.com}

\author{Meng Wang}
\affiliation{
\department[0]{Key Laboratory of Knowledge Engineering with Big Data,}
\institution{Hefei University of Technology}
\institution{Institute of Artificial Intelligence, Hefei Comprehensive National Science Center}
\country{}
}
\email{eric.mengwang@gmail.com}

\author{Yong Li}
\affiliation{
\institution{Beijing National Research Center for Information Science and Technology, Department of Electronic Engineering, Tsinghua University}
\country{}
}
\email{liyong07@tsinghua.edu.cn}

\begin{abstract}
Most modern recommender systems predict users' preferences with two components: user and item embedding learning, followed by the user-item interaction modeling. By utilizing the auxiliary review information accompanied with user ratings, many of the existing review-based recommendation models enriched user/item embedding learning ability with historical reviews or better modeled user-item interactions with the help of available user-item target reviews. Though significant progress has been made, we argue that current solutions for review-based recommendation suffer from two drawbacks. 
First, as review-based recommendation can be naturally formed as a user-item bipartite graph with edge features from corresponding user-item reviews, how to better exploit this unique graph structure for recommendation? Second, while most current models suffer from limited user behaviors, can we exploit the unique self-supervised signals in the review-aware graph to guide two recommendation components better?
To this end, in this paper, we propose a novel \fullname~(\shortname) framework for review-based recommendation. 
Specifically, we first construct a review-aware user-item graph with feature-enhanced edges from reviews, where each edge feature is composed of both the user-item rating and the corresponding review semantics. 
This graph with feature-enhanced edges can help attentively learn each neighbor node weight for user and item representation learning. 
After that, we design two additional contrastive learning tasks~(i.e., Node Discrimination and Edge Discrimination) to provide self-supervised signals for the two components in recommendation process. Finally, extensive experiments over five benchmark datasets demonstrate the superiority of our proposed \shortname~ compared to the state-of-the-art baselines.
\end{abstract}


%

\maketitle

\section{Introduction}
\label{section-1}

As a widely adopted recommendation technique, review-based recommendation tries to model users' preferences to items with the consideration of corresponding reviews~\cite{WSDM2017DeepCoNN, WWW2018NARRE, TOIS2019CARL}. 
Since textual reviews contain more detailed user opinions and item attributes than numerical ratings, they have been demonstrated to be beneficial to mitigate the data sparsity and the cold-start issue \cite{RecSys2013HFT, WSDM2017DeepCoNN, TKDE2021NeuACF++}, thus have drawn much more attention~\cite{WWW2018NARRE, KDD2019DAML, AAAI2020AHN} in recommender systems.

Generally speaking, modern recommender systems usually employ two main components to predict users' preferences: user and item embedding learning, followed by the user-item interaction modeling~\cite{SIGIR2020LightGCN, AAAI2020LR-GCCF, WWW2017NCF, TKDE2022A_Survey_Accuracy-oriented_Recommendation}. 
Compared to users' rating records, the utilization of the auxiliary textual reviews in review-based recommendation can also be categorized into these two classes: \textit{1) Better user/item representation learning by aggregating user~(item) reviews}, \textit{2) User-item interaction modeling with each user-item review record}. 
For the former, historical reviews are usually treated as auxiliary description information for user and item embedding learning~\cite{WSDM2017DeepCoNN, WWW2018NARRE, KDD2019DAML, AAAI2020AHN}. Then, the embeddings are fed into a matching network, such as Factorization Machine~(FM)~\cite{WSDM2017DeepCoNN}, to predict target ratings.
For example, DeepCoNN~\cite{WSDM2017DeepCoNN} concatenates all historical reviews as the descriptive document for user~(item). 
Then, two convolutional neural networks are employed to infer user and item embeddings from their documents. 
Finally, FM is used to predict final ratings based on the learned user and item embeddings.

\begin{figure}[!t]
    \centering
    \includegraphics[width=0.8\linewidth]{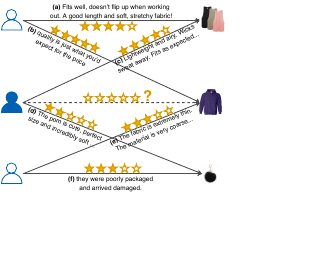}
    \caption{User-item graph with edge features from ratings and reviews.}
    \label{fig:review-based RS}
\end{figure}

However, this historical review usage is too coarse-grained to exploit review information fully. 
As shown in \autoref{fig:review-based RS}, each review describes the specific preference of the user for one specific item~\cite{RecSys2017TransNets} (e.g., review (a) for style and review (b) for price). 
Therefore, a fine-grained utilization of reviews is developed to incorporate the detailed information at the user-item interaction modeling stage. 
To distinguish the historical review usage, we name the type of method as \textit{target review} usage since each review is employed to provide a detailed description for corresponding user-item interaction modeling to predict the \textit{target rating}. 
Unlike historical review usage, target reviews are accessible for observed user-item pairs at the training stage but unavailable at the inference stage. 
Therefore, researchers proposed how to transfer the available review in the training process to facilitate the user-item interaction learning  at the inference stage.
For instance, TransNets~\cite{RecSys2017TransNets} inserts an MLP into DeepCoNN before FM to learn interaction features and forces the learned interaction features to approximate the target review features in Euclidean space. 
SDNets~\cite{TOIS2019SDNet} replaces Euclidean distance with a discriminator network to align interaction features and target review features. 
Then, an adversarial min-max game is applied to reach the alignment through the discriminator. 
Besides, there are other related models exploiting target reviews for better user and item representation learning, such as DualPC~\cite{WWW2020DualPC} and DRRNN~\cite{TNNLS2021DRRNN}.

Despite the great progress they have made, we argue that current solutions for review-based recommendation still have some shortcomings. 
First of all, review-based recommendation naturally forms a user-item bipartite graph with edge features from corresponding user-item reviews. 
How to explore this unique graph structure for better recommendation is still very challenging. 
Second, the user/item embedding learning and user-item interaction modeling in recommendations suffer from limited user behaviors, which suppress the recommendation performance. 
Beyond treating reviews as additional information, is there a better method to develop self-supervised signals (e.g., comparing review representation and user-item interaction representation) for better review utilization in embedding learning and interaction modeling in review-based recommendations? 
This is another important challenge that we focus on in this paper.

To this end, in this paper, we propose \fullname~(\shortname),  a novel graph-based contrastive learning framework that fully explores review information for better recommendations.
Different from existing graph-based recommendation methods that only employ reviews as supplementation for user/item embedding learning and user-item interaction modeling, 
we design a novel \fusingfullname~(\fusingshortname) module to integrate reviews into graph learning, where reviews are used to determine the impact of neighbor nodes on the central node in the user-item graph, as well as calculating the impact of reviews to the central node. 
Along this line, \fusingshortname~can effectively exploit this unique structure of user-item graph with edge features from reviews. 
Moreover, in order to further exploit this unique structure and make full use of review information in edges, we incorporate Contrastive Learning~(CL) to develop self-supervised signals for boosting embedding learning and interaction modeling process based on reviews. 
Specifically, we design Node Discrimination~(ND) and Edge Discrimination~(ED) tasks as additional self-supervised optimization targets. 
ND helps \shortname~ achieve better node embedding learning by comparing the representations in different sub-graphs.   
Moreover ED requires \shortname~to compare the user-item interaction and corresponding review representation, so that the user-item interaction can be better modeled. 
Along this line, \shortname~is able to fully explore review information for better embedding learning and interaction modeling, which is in favor of item recommendations. 
Extensive experiments and detailed analysis over five datasets demonstrate the effectiveness and superiority of our proposed method compared with state-of-the-art review-based baselines.  
The main contributions of this work can be summarized as follows:

\begin{itemize}
    \item We define that review-based recommendation can form a user-item graph with edge features from reviews and propose a novel \fusingshortname~to exploit this unique graph structure for better review utilization and graph learning.
    \item We design a novel \shortname~based on contrastive learning, in which ED and ND tasks are employed to use self-supervised signals for the better user (item) embedding learning and user-item interaction modeling based on reviews in edges in the user-item graph. 
    \item We conduct extensive experiments on five real-world datasets to demonstrate the effectiveness of our proposed \shortname. 
\end{itemize}

\section{Related Work}


\subsection{Review-based Recommendation}


Given the popular two steps of recommender systems: user/item embedding learning and user-item interaction modeling, review-based recommendation models can also be categorized into \textit{Historical reviews utilization for better embedding learning} and \textit{Target review utilization for better interaction modeling}.

\subsubsection{Historical Reviews Utilization}
For example, topic models, such as Latent Dirichlet Allocation (LDA)~\cite{JMLR2003LDA}, can help researchers obtain latent topics in historical reviews. 
Thus, topic information can be used to assist the user and item embedding learning. 
For instance, Collaborative Topic Regression~(CTR)~\cite{KDD2011CTR} used the sum of topic factors and free embeddings to represent items. 
Topic Initialized latent factors Model~(TIM)~\cite{RecSys2020TIM} utilized topic factors to initialize both user and item embeddings. 
Though these methods have shown the usefulness of textual reviews, LDA technique suffers from insufficient text modeling capacity compared to the modern natural language processing techniques. 

To further explore the potential of reviews for better user and item embedding learning, many advanced text methods are introduced into recommendations~\cite{WSDM2017DeepCoNN, RecSys2017D-Attn, TOIS2019CARL, AAAI2020AHN, TOIS-Aspect}. 
For example, Deep Cooperative Neural Networks (DeepCoNN)~\cite{WSDM2017DeepCoNN} used two parallel TextCNNs~\cite{EMNLP2014TextCNN} to extract semantic features from user and item historical reviews and achieved impressive performance. 
To utilize word correlations and review correlations, Paragraph Vector Matrix Factorization (ParVecMF) \cite{WWW2019ParVecMF} employed paragraph vector~ \cite{ICML2014ParagraphVector} to generate better user and item representations from historical reviews. 
Moreover, attention mechanism~\cite{vaswani2017attention, LIANG2020432} is introduced to improve recommendation performance due to its capability of finding the key elements~\cite{zhang2019drr}.
For example, 
Dual attention-based model (D-Attn) is designed to leverage local and global attention layers to select informative words by calculating corresponding attention scores\cite{RecSys2017D-Attn}.
And \citeauthor{WWW2018NARRE}\cite{WWW2018NARRE} proposed NARRE to select relevant reviews in learning the user and item representations with attention mechanism, which can also improve the model interpretability.
Besides, the structure of historical reviews also contains useful information. 
\citeauthor{NAACL2019HUITA}\cite{NAACL2019HUITA} devised a three-tier attention mechanism to make full use of words, sentences, and structure information in reviews. 
There are other relevant methods that explore historical reviews for better user and item embedding learning, such as CARL~\cite{TOIS2019CARL}, DAML~\cite{KDD2019DAML}, and AHN~\cite{AAAI2020AHN}.

Since Graph Convolution Network~(GCN)~\cite{ICLR2017GCN, WWW2021FairGO} can model the natural user-item bipartite graph and has achieved promising results in item recommendations, 
researchers also proposed introducing review signals into the learning of graph-based methods. 
RMG~\cite{EMNLP-IJCNLP2019RMG} is one of the first few attempts to model user preferences from both review and graph perspectives. 
It used hierarchical attention networks to extract review features and attention-based GCN to obtain embeddings for nodes (users and items).
The prediction is calculated based on the concatenation of learned representations based on review features and node embeddings. 
Moreover, \citeauthor{CIKM2020SSG}\cite{CIKM2020SSG} designed a novel Set-Sequence-Graph~(SSG) network to jointly model the multi-view historical reviews, user-item sequences, and the user-item bipartite graph for better user and item embedding learning. 

We also borrow the natural user-item graph structure for review-based recommendation, and we advance these related works by introducing review information and constructing a user-item graph with edge features from corresponding reviews for better user preference modeling and item recommendation.



\subsubsection{Target Reviews Utilization}
Apart from using historical reviews for better embedding learning, incorporating reviews into the user-item interaction modeling stage also attracts much attention. 
Since it leverages review information to boost the corresponding user-item interaction model in a detailed manner, we name this type of method as target review utilization. 
Different from historical review utilization, target reviews are only accessible for observed user-item pairs at the training stage but unavailable at the inference stage. 
To overcome this problem,
\citeauthor{RecSys2017TransNets}\cite{RecSys2017TransNets} designed a novel TransNet to make full use of learned information to approximate the target review so as to make a better recommendation.
Specifically, they proposed a transform layer to approximate target review features based on learned user-item interactions at the training stage.
Euclidean distance is used to constrain the similarity between target reviews and user-item interaction vectors.
Some relevant works are also proposed following this approximation strategy, such as DualPC \cite{WWW2020DualPC} and DRRNN \cite{TNNLS2021DRRNN}. 
Besides, some researchers proposed generation-based methods to tackle the problem of unavailable target reviews at the inference stage. 
For example, SDNet~\cite{TOIS2019SDNet} focused on the consistency of target reviews and user-item interactions. 
It employed GAN to estimate the divergence between target reviews and learned user-item interaction representations so that the target review can be better generated for final prediction.

Despite the great progress they have achieved, current review-based methods only treat reviews as additional information and still suffer from limited user behaviors. 
Different from these methods, we design two CL tasks to develop self-supervised signals to alleviate the limited user behavior problem and achieve better user preference modeling as well as item recommendation.

\subsection{Contrastive Learning}
As one of the representative technologies for self-supervised learning, Contrastive Learning~(CL) has made great progress in Computer Vision~\cite{DIM2018ICLR, InfoNCE} and Natural Language Processing~ \cite{NIPS2013Word2Vec, WWW2019ParVecMF}.
The central intuition of contrastive learning is to pull together an anchor and a “positive” sample in embedding space and push apart the anchor from many “negative” samples~\cite{MINE2018ICML}. 
In recent years, this idea has attracted researchers to explore CL for graph representation learning \cite{DGI2018ICLR, ICML2020GraphMultiview, NIPS2020GraphCL, ICMLW2020GRACE}. 
For instance, \citeauthor{DGI2018ICLR}\cite{DGI2018ICLR} proposed Deep Graph Infomax~(DGI) by maximizing the agreement between node representation and graph representation in a local-global contrastive paradigm. 
\citeauthor{ICMLW2020GRACE}\cite{ICMLW2020GRACE} developed GRACE to generate node representations by maximizing node-level agreement from edge removing and node feature masking augmentations. 
\citeauthor{NIPS2020GraphCL}\cite{NIPS2020GraphCL} designed GraphCL to explore different types of graph augmentations~(such as nodes dropping and edges removing) and analyzed the impacts of various augmentation combinations. 

When it comes to graph-based recommendations, researchers designed various CL paradigms at the user-item interaction modeling stage for performance improvement. 
For example, \citeauthor{WSDM2021BiGI}\cite{WSDM2021BiGI} considered community structures and designed BiGI to recognize the global properties of the bipartite graph. 
\citeauthor{SIGIR2021SGL}\cite{SIGIR2021SGL} unified graph CL tasks and the recommendation task to enhance user and item representation learning. 
There also exist other related works \cite{SIGIR2021EGLN, social2021KDD}. 
They all have achieved impressive performance on graph-based recommendations, which demonstrate the superiority of CL in graph representation learning. 
While most of these works focus on designing graph-based self-supervised learning in collaborative filtering, we utilize the unique self-supervised signals in the review-aware graph for recommendation.

\begin{figure*}[!t]
     \centering
     \includegraphics[width=0.85\linewidth]{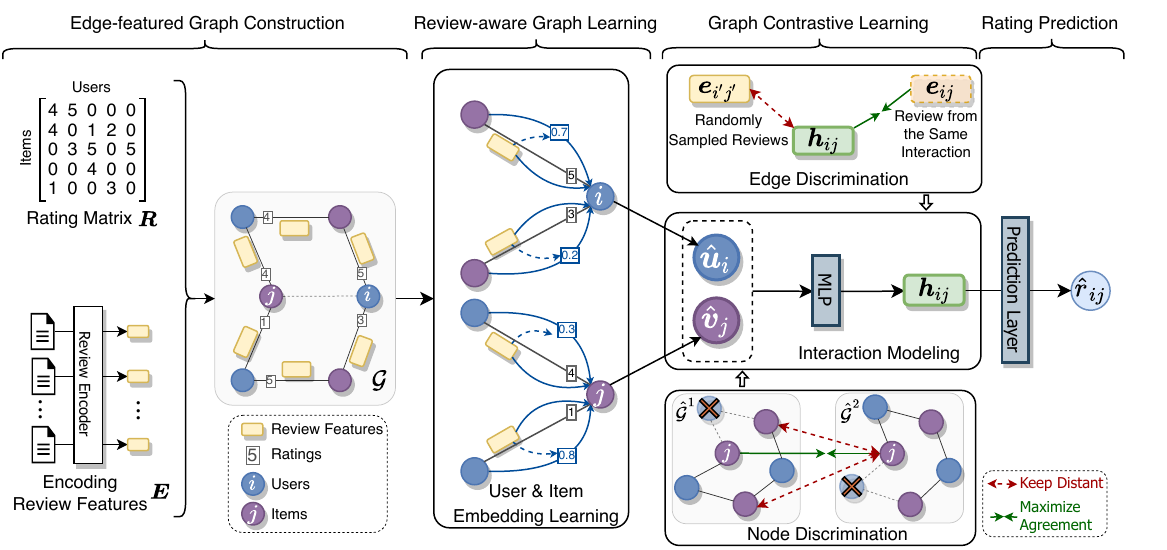}
     \caption{The architecture of \fullname framework.}
     \label{fig:framework}
\end{figure*}

\section{Problem Definition}
\label{s:preliminary}
In a review-based recommendation system, there are two sets of entities: a user set $\mathcal{U}(|\mathcal{U}|= M)$ and an item set $\mathcal{V}(|\mathcal{V}|=N)$, where $u_i\in \mathcal{U}$ and $v_j\in \mathcal{V}$ denote the user $i$ and item $j$. 
$M$ and $N$ denote the number of users and items separately. 
The rating record is represented as a user-item rating matrix $\bm{R} \in \mathcal{R}^{M \times N}$, where each element $r_{ij}$ represents the rating score of user $i$ to item $j$, and $\mathcal{R}$ is the set of all ratings ~(e.g., $\mathcal{R} = \{1, 2, 3, 4, 5\}$ in Amazon).
Meanwhile, we leverage a pre-trained model to process the review that user $i$ commented to item $j$ and obtain a fixed-size vector $\bm{e}_{ij}\in \mathbb{R}^{d}$ as the representation. 
Then, review features of all interactions could be formulated as a tensor $\bm{E} \in \mathbb{R}^{M \times N \times d}$. 
The rating matrix $\bm{R}$ and review tensor $\bm{E}$ make up the user-item interactions $ \mathcal{E} = \{\bm{R}, \bm{E}\}$. 
To this end, the review-based recommendation data can be formulated as a bipartite graph with featured edges $\mathcal{G}=<\mathcal{U} \cup \mathcal{V}, \mathcal{E}>$, with each observed edge containing both the detailed rating value and the semantic vector obtained from the corresponding review. The task of review-based recommendation is to predict the final rating matrix $\hat{\bm{R}}\in\mathcal{R}^{M\times N}$ with the graph $\mathcal{G}$.

\section{\fullname}
\label{s:model}
Figure~\ref{fig:framework} illustrates the overall architecture of our proposed \shortname, which consists of two main modules: 
1) \textit{Review-aware Graph Learning Module}: incorporating the reviews into user preference modeling based on graph input;
2) \textit{Graph Contrastive Learning Module}: introducing CL technology to promote recommendation performance from two perspectives: user/item embedding learning and interaction modeling. In the following, we will give a detailed explanation of the two modules. 



\subsection{Review-aware Graph Learning Module}
As mentioned before, we aim at fully exploiting review information to learn accurate user preferences. Specifically, we design a review-aware graph learning module consisting of two key components in recommendation systems: embedding learning and interaction modeling. The module has two advances as follows. Firstly, it naturally formulates the available data as a user-item bipartite graph with ratings and reviews. Besides, we construct a review-aware graph learning that could better capture the correlation between collaborative signals and reviews. Next, we introduce each component in detail.

\subsubsection{Review-aware User and Item Embedding Learning}
There are three components: 
\textit{Initialization}: initializing all the inputs, including node embeddings and edge features; 
\textit{Review-aware Message Passing}: leveraging rating and review information on edges to measure the influence of neighbor nodes and historical reviews to the central node; 
\textit{Message Aggregation}: aggregating messages from neighbors for the central node embedding learning. 
The details of each component are listed in the following parts.



\textbf{Initialization.} 
In this component, we leverage free embedding matrices $\bm{U}\in\mathbb{R}^{M\times d}$ and $\bm{V}\in\mathbb{R}^{N\times d}$ to denote users and items. 
$\boldsymbol{u}_i \in\mathbb{R}^{d}$ and $\boldsymbol{v}_j \in\mathbb{R}^{d}$ denote free embeddings of the user $i$ and item $j$, specifically. 
Following \cite{2017GC-MC}, we treat rating score as edge type to distinguish the semantics carried by ratings.
For review representations, similar to~\cite{SIGIR2018SentiRec}, we employ  BERT-Whitening~\cite{su2021whitening} to generate the representation $\bm{e}_{ij} \in\mathbb{R}^{d}$ for the review that user $i$ commented on item $j$
\footnote{Compared with concatenation of raw review text, pre-encoded reviews that embody the corresponding ratings have been proved to reduce the training time and the memory usage without decreasing recommendation performance \cite{SIGIR2018SentiRec}.}.
We have to note that the review representation will be frozen during the model training to reduce the training time and memory usage. 

\textbf{Review-aware Message Passing.}
As mentioned above, each interaction contains a rating and a review semantic feature.
Compared to the numerical rating, textual reviews contain fine-grained semantics, which has following advantages. First, the review semantics contain detailed user preferences and item attributes that can help learn better user and item embeddings \cite{WSDM2017DeepCoNN, WWW2018NARRE}. Second, the detailed review semantics can help the model learn the precise extent users like/dislike the items, which can be further employed to re-weight the impacts between users and items.
Thus, as shown in Figure~\ref{fig:framework}, we utilize review features to fine-tune the influences of neighbor $j$ and review itself on the central node $i$, which can be formulated as the rating-specific message passing:
\begin{equation}
    \label{eq:RGC}
        \boldsymbol{x}^{(l)}_{r; j \rightarrow i} = \frac{
        \sigma (\boldsymbol{w}^{(l){\top}}_{r,1}\boldsymbol{e}_{ij}) \boldsymbol{W}^{(l)}_{r,1} \boldsymbol{e}_{ij} 
        + \sigma (\boldsymbol{w}^{(l){\top}}_{r,2}\boldsymbol{e}_{ij}) \boldsymbol{W}^{(l)}_{r,2} \boldsymbol{v}^{(l-1)}_{j} 
        }{\sqrt{|\mathcal{N}_{j}||\mathcal{N}_{i}|}},
\end{equation}
where $\bm{e}_{ij}$ is the review representation that user $i$ commented on item $j$. 
$\bm{v}^{(l-1)}_j$ denotes the embedding of item $j$ learned from the $l-1$ layer, where $\bm{v}^{(0)}_j$ is initialized with free embedding $\bm{v}_{j}$.
$\{\boldsymbol{w}^{l}_{r,1}, \boldsymbol{w}^{l}_{r,2}, \boldsymbol{W}^{l}_{r,1}, \boldsymbol{W}^{l}_{r,2} \mid r \in \mathcal{R}\} $ are rating-specific trainable parameters at the $l$-th propagation layer based on rating $r$.  
Among them, $\boldsymbol{W}^{l}_{r,1}$ and $ \boldsymbol{W}^{l}_{r,2}$ are used to transform the review representation $\boldsymbol{e}_{ij}$ and neighbor node embedding $\boldsymbol{v}_j$ to the same space. 
$\boldsymbol{w}^{l}_{r,1}$ and $\boldsymbol{w}^{l}_{r,2}$ are used to learn two scalar weights from the review feature $\bm{e}_{ij}$ to re-weight the impacts of the neighbor node and review itself on the central node.
$\sigma(\cdot)$ is the sigmoid activation function. 
$\mathcal{N}_i$ and $\mathcal{N}_j$ represent the set of items rated by user $i$ and the set of users who rated item $j$.
Meanwhile, we can obtain the message passing from user $i$ to item $j$ in an analogous way:
\begin{equation}
    \label{eq:RGC1}
        \boldsymbol{x}^{(l)}_{r; i \rightarrow j} = \frac{
        \sigma (\boldsymbol{w}^{(l){\top}}_{r,1}\boldsymbol{e}_{ij}) \boldsymbol{W}^{(l)}_{r} \boldsymbol{e}_{ij} 
        + \sigma (\boldsymbol{w}^{(l){\top}}_{r,2}\boldsymbol{e}_{ij}) \boldsymbol{W}^{(l)}_{r,2} \boldsymbol{u}^{(l-1)}_{i} 
        }{\sqrt{|\mathcal{N}_i||\mathcal{N}_j|}},
\end{equation}

\textbf{Message Aggregation.}
Similar to the aggregation operation in GC-MC, we use the following operation to aggregate all messages together to generate user and item embeddings at the $l$-th layer after the message passing operation:
\begin{equation}
\begin{split}
    \label{eq:aggregation}
    \bm{u}^{(l)}_i = \boldsymbol{W}^{(l)} \sum_{r \in \mathcal{R}} \sum_{k \in \mathcal{N}_{i,r}} \boldsymbol{x}^{(l)}_{r; k \rightarrow i} \ , \quad
    \bm{v}^{(l)}_j = \boldsymbol{W}^{(l)} \sum_{r \in \mathcal{R}} \sum_{k \in \mathcal{N}_{j,r}} \boldsymbol{x}^{(l)}_{r; k \rightarrow j} \ ,
\end{split}
\end{equation} 
where $\bm{u}^{(l)}_i$ and $\bm{v}^{(l)}_j$ are the learned embeddings at the $l$-th layer for user $i$ and item $j$. 
$\boldsymbol{W}^{(l)}$ is the trainable parameter matrix.
$\mathcal{N}_{i,r}$ denotes the set of items that rated by user $i$ with rating $r$.
$\mathcal{N}_{j,r}$ is the set of users that rated item $j$ with rating $r$.

By stacking $L$ layers of message passing and aggregation, we take the outputs $\bm{u}^{(L)}_i$ and $\bm{u}^{(L)}_j$ from the $L$-th layer as the final user and item embeddings:

\begin{equation}
    \hat{\bm{u}}_i = \bm{u}^{(L)}_i, \quad \hat{\bm{v}}_j = \bm{v}^{(L)}_j,
\end{equation}


\subsubsection{Interaction Modeling}
To model the interaction from user and item embeddings, we take the outputs $\hat{\bm{u}}_i$ and $\hat{\bm{v}}_j$ as the input, and leverage an Multi-Layer Perceptron (MLP) to obtain corresponding interaction features $\bm{h}_{ij}$, which can be formulated as follows:
\begin{equation}
    \label{eq:interaction feature}
        \boldsymbol{h}_{ij} = \operatorname{MLP} \left(\left[\hat{\bm{u}}_i, \hat{\bm{v}}_j\right]\right),
\end{equation}
where $\boldsymbol{h}_{ij} \in \mathbb{R}^d$ is the learned user-item interaction features. $\operatorname{MLP}(\cdot)$ denotes the MLP with two hidden layers and GELU activation function. 
$[, ]$ denotes the concatenation operation. After getting $\boldsymbol{h}_{ij}$, we can predict the rating based on the interaction features:
\begin{equation}
    \label{eq:decoder}
        \hat{r}_{ij} = \boldsymbol{w}^{\top} \bm{h}_{ij}, \\
\end{equation}
where $\boldsymbol{w}$ is a parameter vector that map interaction features to the prediction ratings.

\subsection{Contrastive Learning Module}
In the above parts, our proposed \fusingshortname~ takes both advantages of textual reviews and graph signals for user and item representations. 
Following the supervised learning framework \cite{WWW2018NARRE, 2017GC-MC}, we can optimize  parameters in \fusingshortname~ by encouraging the predicted ratings to be close to the observed ratings.
However, the limited interaction behaviors restricted the accurate user preference learning, leading to recommender models being far from satisfactory.
To alleviate this shortcoming, we intend to leverage Contrastive Learning~(CL) to develop self-supervised signals to boost \fusingshortname~learning.
Specifically, we design two CL paradigms, i.e., \textit{Node Discrimination} (ND) to enhance node embedding learning and \textit{Edge Discrimination} (ED) to boost interaction modeling. 

\subsubsection{Node-based CL for Node Representation Enhancement}
Following previous work that applies the CL to graph learning~\cite{ICML2020GraphMultiview, NIPS2020GraphCL, ICMLW2020GRACE}, we also utilize the CL paradigm to promote better learning of graph structure and the node (i.e., users and items) embedding learning.
Specifically, we employ node dropping to generate sub-graphs for contrastive learning. 
Taking the user node embedding learning as an example, we randomly discard item nodes and corresponding review features with probability $\rho$ to generate sub-graphs $\hat{\mathcal{G}}^1$ and $\hat{\mathcal{G}}^2$ for message passing and aggregation. 
By applying our proposed \fusingshortname~to these two sub-graphs, we can obtain the two different embeddings $\hat{\bm{u}}_i^1$ and $\hat{\bm{u}}_i^2$ for user $i$, which can be treated as the positive pair. 
Similarly, we can obtain the negative node embedding $\hat{\bm{u}}_{i'}^2$ of different user $i'$ from sub-graph $\hat{\mathcal{G}}^2$, where $i' \neq i$. 
Therefore, the target is maximizing the similarity of the positive pair $(\hat{\bm{u}}_i^1, \hat{\bm{u}}_i^2)$ and minimizing the similarity of the negative pair $(\hat{\bm{u}}_i^1, \hat{\bm{u}}_{i'}^2)$. 
This process can be formulated as follows: 

\begin{equation}
\mathcal{L}^{user}_{nd}
=
- \mathbb{E}_{\mathcal{U}} \left[\log\left(F\left(\hat{\bm{u}}_i^1, \hat{\bm{u}}_i^2 \right)\right) \right]
+ \mathbb{E}_{\mathcal{U} \times \mathcal{U}'} \left[\log\left(F\left(\hat{\bm{u}}_i^1, \hat{\bm{u}}_{i'}^2 \right)\right) \right],
\label{eq:MI:user}
\end{equation}
where $i$ is the input user, $i'$ is the negative user sampled from $\mathcal{U}' = \mathcal{U}$. 
And $F\left(\hat{\bm{u}}_i^1, \hat{\bm{u}}_i^2 \right)=\sigma\left(\hat{\bm{u}}_i^{1\top} \bm{W} \hat{\bm{u}}_i^2 \right)$ is the similarity function with trainable parameter $\bm{W}$ and sigmoid activation function $\sigma$.
Analogously, we can obtain the optimization target on item nodes $\mathcal{L}^{\textrm{item}}_{nd}$. 
Combining the ND tasks on both user and item nodes, we get the final ND target:  $\mathcal{L}_{nd} = \mathcal{L}^{user}_{nd} +  \mathcal{L}^{item}_{nd}$.

\subsubsection{Edge-based CL for Target Review Alignment}

The review on each edge carries a more detailed assessment from the user to the item than the numerical ratings. 
To exploit the effective reviews for better interaction modeling, we propose to leverage Edge Discrimination (ED) as the additional CL task to achieve this goal.

After obtaining the interaction feature $\bm{h}_{ij}$ from \autoref{eq:interaction feature}, we treat it as the anchor example. 
Following the CL framework~\cite{chen2020simple}, we select the corresponding review representation $\bm{e}_{ij}$ as the positive sample. 
Naturally, a randomly sampled review representation $\bm{e}_{i'j'}$ from the entire training set is treated as the negative sample. 
We have to note that the conditional sampled method, such as selecting the other reviews that user $i$ commented on or item $j$ received, makes no difference with the randomly sampling. 
The target of ED is to pull together anchor example $\bm{h}_{ij}$ and positive example $\bm{e}_{ij}$, as well as push away anchor example $\bm{h}_{ij}$ and negative example $\bm{e}_{i'j'}$. 
Thus, the optimization target can be formulated as follows:
\begin{equation}
\mathcal{L}_{ed}
=
- \mathbb{E}_{\mathcal{E}}\left[\log\left(F\left(\bm{h}_{ij}, \bm{e}_{ij} \right)\right)\right] 
+ \mathbb{E}_{\mathcal{E} \times \mathcal{E}'}\left[\log\left(F\left(\bm{h}_{ij}, \bm{e}_{i'j'} \right)\right)\right],
\label{eq:MI:edge}
\end{equation}
where $\bm{e}_{ij}$ is the target review feature corresponding to the interaction $(i, j)$.  
And $\bm{e}_{i'j'}$ is the negative review feature corresponding to the interaction $(i',j')$ sampled from $\mathcal{E}'=\mathcal{E}$. 

\subsection{Model Optimization}
Since \shortname~focuses on predicting the ratings of users to items, we employ Mean Square Error (MSE) as the optimization target, which is widely adopted in current rating prediction works~\cite{NIPS2007PMF, WSDM2017DeepCoNN}: 
\begin{equation}
    \label{eq:main loss}
    \mathcal{L}_{main} = \frac{1}{|\mathcal{S}|} \sum_{(i,j) \in \mathcal{S}} (\hat{r}_{ij} - r_{ij})^2,
\end{equation}
where $\mathcal{S}$ represents user-item pairs in the training set. $r_{ij}$ is the observed rating that user $i$ commented on item $j$. 
Meanwhile, we employ CL to constrain the learning process. 
We optimize the recommendation and CL tasks simultaneously with the importance hyper-parameters $\alpha$ and $\beta$:
\begin{equation} \label{eq:total loss}
\mathcal{L} = 
\mathcal{L}_{main} + \alpha\mathcal{L}_{ed} + \beta\mathcal{L}_{nd}.
\end{equation}

\section{Experiments}

\begin{table}[t]
\caption{Statistics of datasets.}
\label{tab:Dataset_Statistics}
    \begin{tabular}{crrrrr}
    \toprule
    Datasets         & \#Users  & \#Items  & \#Reviews   & Density  \\
    \midrule 
    Digital\_Music   & 5,541    & 3,568   & 64,706     & 0.330\%  \\
    Toys\_and\_Games & 19,412   & 11,924  & 167,597    & 0.072\%  \\
    Clothing         & 39,387   & 23,033  & 278,677    & 0.031\%  \\
    CDs\_and\_Vinly  & 75,258   & 64,443  & 1,097,592  & 0.023\%  \\
    Yelp             & 8,423    & 3,742   & 88,647     & 0.281\%  \\
    \bottomrule
    \end{tabular}
\end{table}

\subsection{Experimental Settings}
\subsubsection{Datasets. }
We evaluate our model on five benchmark datasets. 
Four of them belong to Amazon  5-core~\cite{AmazonData2019}\footnote{http://jmcauley.ucsd.edu/data/amazon/links.html} in different domains: ``Digital Music'', ``Toys and Games'', ``Clothing'', and ``CDs and Vinyl''. 
The 5-core means there are at least five reviews for each user or item.
Following \cite{RecSys2017D-Attn}, we also conduct experiments on Yelp Business Rating Prediction Challenge 2013 dataset\footnote{https://www.kaggle.com/c/yelp-recsys-2013}, which contains restaurant reviews in Phoenix, AZ metropolitan area. 
We have pre-processed the raw data of Yelp following the 5-core setting. The rating scores of these five datasets are integers from one to five. 
Following similar works~\cite{RecSys2017TransNets, KDD2019DAML}, each dataset is randomly split into training, validation, and test sets with 80\%, 10\%, 10\%, respectively. 
The statistics of these five datasets are summarized in \autoref{tab:Dataset_Statistics}.

\subsubsection{Evaluation Metric. } 
Following \cite{TOIS2019CARL, TOIS2019SDNet}, the performance is evaluated by \textit{MSE}, which is widely used for rating prediction in recommendation systems. 
For fair comparison, we repeat each experiment \textbf{five times} and report \textit{mean(±std)} for model comparison. 
For review-based recommendation, a relative improvement above 1\% is regarded as significant~\cite{KDD2018MPCN, EMNLP2021ASPE}.

\begin{table*}[!t]
  \begin{threeparttable}
    \caption{Results in terms of the MSE on five datasets of different methods.}
    \label{tab:overall_performance}
    \begin{tabular}{lcccccc}
    \hline \hline
    Method         & Digital\_Music                     & Toys\_and\_Games                 & Clothing                           & CDs\_and\_Vinly                   & Yelp   \\ \hline \hline
    (1) SVD            & 0.8523\small{$\pm$4e-4}            & 0.8086\small{$\pm$1e-3}          & 1.1167\small{$\pm$1e-3}            & 0.8662\small{$\pm$2e-4}           & 1.1939\small{$\pm$1e-3} \\
    (2) NCF            & 0.8403\small{$\pm$5e-3}            & 0.8078\small{$\pm$2e-3}          & 1.1094\small{$\pm$1e-3}            & 0.8781\small{$\pm$1e-3}           & 1.1896\small{$\pm$4e-3} \\ \hline
    (3) DeepCoNN       & 0.8378\small{$\pm$1e-3}            & 0.8028\small{$\pm$7e-4}          & 1.1184\small{$\pm$2e-3}            & 0.8621\small{$\pm$1e-3}           & 1.1877\small{$\pm$1e-3} \\
    (4) NARRE          & 0.8172\small{$\pm$1e-3}            & 0.7962\small{$\pm$1e-3}          & 1.1064\small{$\pm$1e-3}            & 0.8495\small{$\pm$1e-3}           & 1.1862\small{$\pm$1e-3} \\
    (5) DAML           & 0.8237\small{$\pm$2e-3}            & 0.7936\small{$\pm$4e-3}          & 1.1065\small{$\pm$2e-3}            & 0.8483\small{$\pm$1e-3}           & 1.1793\small{$\pm$1e-3} \\ \hline
    (6) SDNet          & 0.8331\small{$\pm$3e-3}            & 0.8006\small{$\pm$1e-3}          & 1.1080\small{$\pm$1e-3}            & 0.8654\small{$\pm$5e-4}           & 1.1837\small{$\pm$3e-3} \\
    (7) TransNets      & 0.8273\small{$\pm$5e-3}            & 0.7980\small{$\pm$1e-2}          & 1.1141\small{$\pm$5e-3}            & 0.8440\small{$\pm$1e-3}           & 1.1855\small{$\pm$2e-3} \\ \hline 
    (8) GC-MC          & 0.8090\small{$\pm$1e-3}            & 0.7986\small{$\pm$5e-4}          & 1.1088\small{$\pm$1e-3}            & \underline{0.8404}\small{$\pm$1e-3}           & 1.1737\small{$\pm$1e-3} \\ 
    (9) RMG            & \underline{0.8074}\small{$\pm$1e-3}            & \underline{0.7901}\small{$\pm$1e-3}          & \underline{1.1064}\small{$\pm$2e-3}            & 0.8425\small{$\pm$8e-4}           & \underline{1.1705}\small{$\pm$1e-3} \\
    (10) SSG            & 0.8218\small{$\pm$2e-3}            & 0.8064\small{$\pm$1e-3}          & 1.1228\small{$\pm$1e-3}            & 0.8458\small{$\pm$1e-3}           & 1.1807\small{$\pm$1e-3} \\ \hline
    (11) \fusingshortname            & 0.8037\small{$\pm$2e-3} (0.5\%)    & 0.7853\small{$\pm$8e-4} (0.6\%)  & 1.1024\small{$\pm$9e-4} (0.4\%)    & 0.8360\small{$\pm$1e-3} (0.5\%)   & 1.1692\small{$\pm$2e-3} (0.1\%) \\
    (12) \fusingshortname+ND         & 0.7780\small{$\pm$2e-3} (3.6\%)    & 0.7831\small{$\pm$1e-3} (0.9\%)  & 1.0925\small{$\pm$3e-4} (1.3\%)    & 0.8240\small{$\pm$6e-4} (2.0\%)   & 1.1625\small{$\pm$1e-3} (0.7\%) \\ 
    (13) \fusingshortname+ED         & 0.7810\small{$\pm$3e-3} (3.3\%)    & 0.7797\small{$\pm$9e-4} (1.3\%)  & 1.0891\small{$\pm$8e-4} (1.6\%)    & 0.8244\small{$\pm$1e-3} (1.9\%)   & 1.1636\small{$\pm$1e-3} (0.6\%) \\
    (14) \shortname      & \textbf{0.7735{$\pm$4e-3} (4.2\%)}    & \textbf{0.7771{$\pm$1e-4} (1.6\%)} & \textbf{1.0858{$\pm$1e-3} (1.9\%)}  & \textbf{0.8180{$\pm$7e-4} (2.7\%)} & \textbf{1.1609{$\pm$8e-4} (0.8\%)} \\ \hline \hline
    \end{tabular}
    \begin{tablenotes}
      \small
      \item The best results are highlighted in bold. The percentages indicate the relative improvements over the best baselines marked by underline. All the results are reported as “mean(±std)” across 5 random runs.
    \end{tablenotes}
  \end{threeparttable}
\end{table*}

\subsubsection{Baselines. }
We select conventionally and recently published review-based baselines for model comparison, including advanced graph-based methods. They are listed as follows:

\begin{itemize}
    
    \item \textbf{SVD}~\cite{Computer2009SVD} is a classical matrix factorization model that estimates ratings by the inner product of users' and items' latent factors.
    \item \textbf{NCF}~\cite{WWW2017NCF} uses a neural network to predict the rating based on user and item free embeddings.
    \item \textbf{DeepCoNN}~\cite{WSDM2017DeepCoNN} is one of the pioneer works that extracts user/item feature from documents (concatenations of reviews) using neural networks. 
    \item \textbf{NARRE}~\cite{WWW2018NARRE} improves upon DeepCoNN by hiring an attention mechanism to estimate the usefulness of different reviews.
    \item \textbf{DAML}~\cite{KDD2019DAML} enhances user and item representation by modeling the interaction between user and item documents.
    \item \textbf{SDNet}~\cite{TOIS2019SDNet} proposes a GAN-based \cite{NIPS2014GAN} distillation method to transform informative target review signal into NCF. 
    \item \textbf{TransNets}~\cite{RecSys2017TransNets}: inserts an MLP into DeepCoNN to transform user and item features to an approximation of target review features.
    \item \textbf{GC-MC}~\cite{2017GC-MC} regards rating prediction as link prediction on the user-item bipartite graph and adopts relational graph convolution~\cite{ESWC2018RGCN} to encode user and item embeddings.
    \item \textbf{RMG}~\cite{EMNLP-IJCNLP2019RMG} is one of the first models that fuse graph signals and review information. 
    \item \textbf{SSG}~\cite{CIKM2020SSG} jointly models review sets, review sequences, and user-item graphs. The authors design the Review-aware graph attention network (RGAT) to capture the graph signals for the user-item graph.
    
\end{itemize}

Above mentioned methods can briefly categorized into five groups:
(1) \textit{Traditional rating-based collaborative filtering methods}, SVD \cite{Computer2009SVD} and NCF \cite{WWW2017NCF};
(2) \textit{Historical review-based methods}: DeepCoNN \cite{WSDM2017DeepCoNN}, NARRE \cite{WWW2018NARRE}, DAML \cite{KDD2019DAML};
(3) \textit{Target review-based methods}: TransNets \cite{RecSys2017TransNets}, SDNet \cite{TOIS2019SDNet};
(4) \textit{Graph-based method}: GC-MC \cite{2017GC-MC}.
(5) \textit{Graph and review fusing methods}: RMG \cite{EMNLP-IJCNLP2019RMG} and SSG \cite{CIKM2020SSG}. 
We have to note that we reimplement RMG, replace the attentive graph neural with GC-MC, and assign separate convolution channels for each rating of RGAT in SSG for a fair comparison.

\subsubsection{Implementation Details. }
For reviews, we leverage BERT-Whitening~\cite{su2021whitening} to encode each review to a fixed-size feature vector, which will not be updated during model training. 
In the final architecture of \fusingshortname~, we utilize one layer message passing.
The size of embeddings (users, items, and reviews) is set as $d=64$.
We have tested the hyper-parameters $\alpha$ in a range of \{0.2, 0.4, 0.6, 0.8, 1.0, 2.0\}, $\beta$ in a range of \{0.2, 0.4, 0.6, 0.8, 1.0\} and node dropout ration in \{0.6, 0.7, 0.8, 0.9\}. 
All the trainable parameters are initialized with the Xavier method, and we select Adam \cite{ICLR2015Adam} as the optimizer for the entire model training. 
The entire model is implemented with Deep Graph Library\footnote{https://www.dgl.ai} and Pytorch\footnote{https://pytorch.org} based on Nvidia RTX GPU.

\subsection{Performance Evaluation}

\subsubsection{Overall Performance Comparison}
\autoref{tab:overall_performance} reports the overall results among five datasets. 
According to the results, we can obtain the following observations:

Firstly, review-based baselines (\autoref{tab:overall_performance}~(3)-(7)) achieve impressive performance, proving the effectiveness of reviews. 
Moreover, target reviews can improve model performance due to their high relevance with users and items.

Secondly, graph-based baselines (\autoref{tab:overall_performance}~(8)-(10)) are able to model complex user-item interactions, thus achieving the best performance over all baselines. This phenomenon supports the importance of graph learning in dealing with higher-order signals. Moreover, we can obtain that the performance of SSG is not as good as the other two graph-based baselines. We speculate the possible reason is that SSG ignores the importance of collaborative filter signals among users and items.

Thirdly, Our proposed \shortname~achieves the best performance across all datasets. Compared with all baselines, \shortname~has two advantages to achieve the best performance. 
First of all, \shortname~utilizes a newly designed \fusingshortname~to integrate the advantages of historical reviews in graph learning, so that the relevant reviews can be fully explored and the noise problem in reviews will be alleviated.   
Moreover, we propose to employ two CL tasks (i.e., ND and ED) for better node embedding and interaction feature learning. These two additional CL tasks can help \fusingshortname~to pay close attention to target review signals and embedding learning for user and item, which is in favor of user preference modeling and final rating prediction.

\begin{figure*}[!t]
    \begin{subfigure}[b]{0.3\textwidth}
        \centering
        \includegraphics[width=\textwidth]{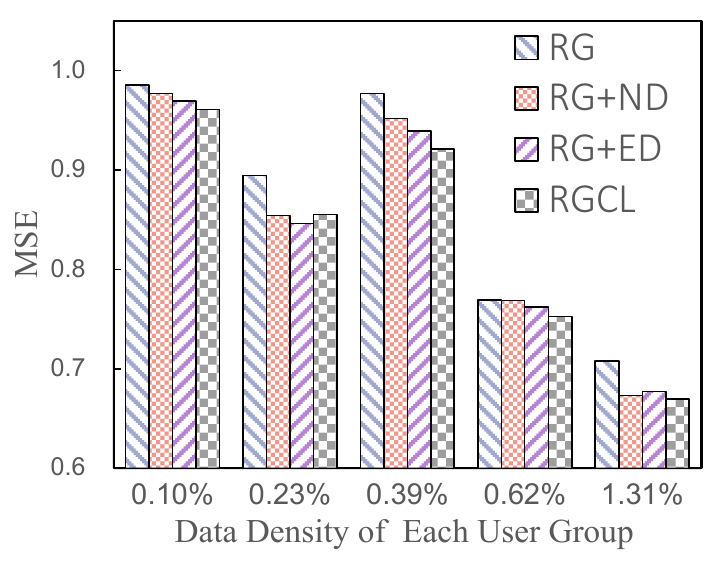}
        \caption{Digital\_Music}
    \end{subfigure}
    \begin{subfigure}[b]{0.3\textwidth}
        \centering
        \includegraphics[width=\textwidth]{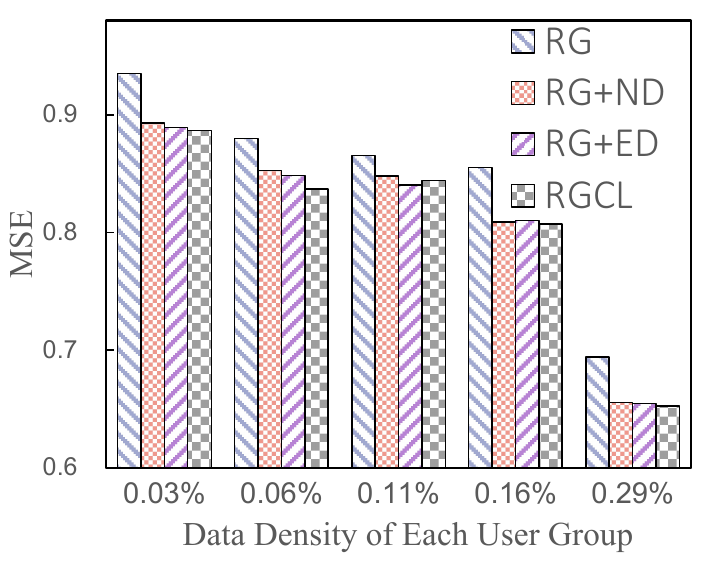}
        \caption{Toys\_and\_Games}
    \end{subfigure}
    \begin{subfigure}[b]{0.3\textwidth}
        \centering
        \includegraphics[width=\textwidth]{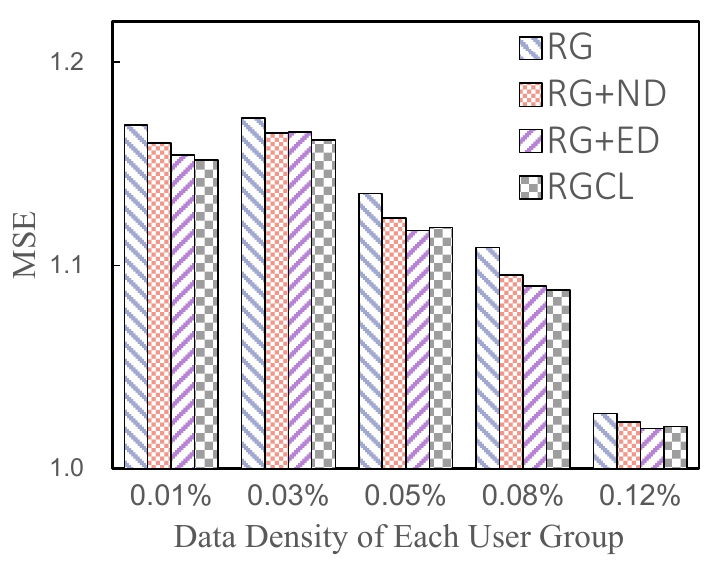}
        \caption{Clothing}
    \end{subfigure}
    
    \caption{Performance comparison over different user groups among \fusingshortname, \fusingshortname+ND, \fusingshortname+ED and \shortname. The percentages on the abscissa represent the density in the group.}
    \label{fig:Perform_Sparsity}
\end{figure*}

\subsubsection{Performance under Different Data Sparsity}
As mentioned in Section~\ref{section-1}, review information can be used to alleviate the data sparsity problem in recommendation. 
The overall experimental results in \autoref{tab:overall_performance} have proven that \shortname~can make full use of reviews to alleviate this problem and provide better recommendations. 
In order to further analyze how \shortname~achieves this goal, we have conducted additional experiments with different sparsity settings. 
For users in training the set, we split them into five groups according to their interaction numbers with items. 
Specifically, we rank the users on each dataset according to their interaction numbers, and then we uniformly split all users into five groups.
Then, we report the MSE comparison of \fusingshortname, \fusingshortname+ND, \fusingshortname+ED, and RGCL in these user groups, as illustrated in \autoref{fig:Perform_Sparsity}. 
According to these results, we have following observations:

Firstly, CL tasks can make better use of reviews and alleviate the data sparsity problem to some extent. Moreover, compared with ND, ED focuses on the consistency of user-item interaction and corresponding review representation, which can achieve better utilization of review information.

Secondly, Our proposed \shortname~leverages \fusingshortname~to integrate the advantages of historical review and target review utilization, as well as employ two CL tasks for better users/items embedding and interaction modeling. Therefore, \shortname~is able to effectively alleviate the data sparsity problem and have the best performance across different data sparsity settings.

\subsection{The Effectiveness of \fusingshortname~}
\label{exp:RGC}
We design a novel \fusingshortname~in \shortname~to better integrate the advantages of historical review utilization and target review utilization. 
To better investigate the effectiveness of \fusingshortname, we conducted an ablation study on different operations~(i.e., weighted review features and the re-weight operation) in \fusingshortname. 
The results are illustrated in \autoref{tab:RGC_Abalation}, where \textit{\fusingshortname~(w/o review)} denotes that reviews are only used to calculate the weight of corresponding neighbors for the central node, and \textit{\fusingshortname~(w/o weight)} denotes that reviews are only used as impact factors of the central node. We list the detailed message passing of \textit{\fusingshortname~(w/o review)} and \textit{\fusingshortname~(w/o weight)} as follows:

$$
\begin{aligned}
    \textit{\fusingshortname(w/o review)}&:  \boldsymbol{x}^{(l)}_{r;j \rightarrow i} = \frac{\sigma (\boldsymbol{w}^{(l){\top}}_{r,2}\boldsymbol{e}_{ij}) \boldsymbol{W}^{(l)}_{r,2} \boldsymbol{v}^{(l-1)}_{j}}{\sqrt{|\mathcal{N}_{j}||\mathcal{N}_{i}|}},  \\
    \textit{\fusingshortname (w/o weight)}&:  \boldsymbol{x}^{(l)}_{r;j \rightarrow i} = 
    \frac{
        \sigma (\boldsymbol{w}^{(l){\top}}_{r,1}\boldsymbol{e}_{ij}) \boldsymbol{W}^{(l)}_{r,1} \boldsymbol{e}_{ij} 
        + \boldsymbol{W}^{(l)}_{r,2} \boldsymbol{v}^{(l-1)}_{j} 
        }{\sqrt{|\mathcal{N}_{j}||\mathcal{N}_{i}|}}.
\end{aligned}
$$

From the results, we can observe that both \textit{\fusingshortname(w/o review)} and \textit{\fusingshortname(w/o weight)} have better performance than GC-MC and RGAT, indicating that our proposed \fusingshortname~can better explore review information. 
Moreover, \textit{\fusingshortname(w/o review)} has better performance than \textit{\fusingshortname(w/o weight)}. It demonstrates that reviews are critical for determining the weights of impacts from different neighbor nodes to the central node. 
The results of \fusingshortname~demonstrate that integrating both operations on reviews can fully explore the advantages of historical review utilization, thus improving the model performance effectively. 

Furthermore, we conduct experiments to investigate the impact of the message passing layer number $L$. In detail, we set the number of layers in {1, 2, 3} and take the user and item embeddings at the last layer as the final embeddings. Then, we report their MSE results in \autoref{tab:RGC_Layers}. 
The MSE results with two layers have a slightly decrease than the performance with one layer.
And MSE with three layers shows a relative lager performance reduction.
\fusingshortname~achieves the smallest MSE with 1 layer propagation. 
We speculate the possible reason is that the review feature propagates multi-layers may bring more irrelevant information to node representations.
Apart from simply treating the output of the $L$-th layer as the final representations of users and items, inspired by LR-GCCF~\cite{AAAI2020LR-GCCF}, we also concatenate the output representations of all $L$ layers of the users and items, then we treat the concatenated representations as users' and items' final representations. And we also find our proposed model only achieves the best performance when $L$ is set to 1.

\begin{table}[t]
\caption{MSE comparison of different components of \fusingshortname}
\resizebox{1\linewidth}{!}{
\begin{tabular}{@{}lccc@{}}
\toprule
Models      & Digital\_Music                    & Toys\_and\_Games              & Clothing                   \\ \midrule
GC-MC       & 0.8145\small{$\pm$1e-3}           & 0.8034\small{$\pm$5e-4}       & 1.1088\small{$\pm$1e-3}    \\ 
RGAT        & 0.8248\small{$\pm$2e-3}           & 0.8093\small{$\pm$3e-3}       & 1.1158\small{$\pm$2e-3}    \\ \hline
\textit{\fusingshortname~\small{(w/o review)}}     & 0.8077\small{$\pm$1e-3}           & 0.7927\small{$\pm$1e-3}       & 1.1038\small{$\pm$1e-3}    \\ 
\textit{\fusingshortname~\small{(w/o weight)}}     & 0.8074\small{$\pm$2e-3}           & 0.7901\small{$\pm$1e-3}       & 1.1064\small{$\pm$1e-3}    \\ 
\fusingshortname~         & 0.8037\small{$\pm$2e-3}           & 0.7853\small{$\pm$8e-4}       & 1.1024\small{$\pm$9e-4}    \\ 
\bottomrule
\end{tabular}}
\label{tab:RGC_Abalation}
\end{table}

\vspace{2mm}
\subsection{The Effectiveness of CL tasks}
Apart from \fusingshortname, we also employ two CL tasks~(i.e., ED and ND) to help \shortname~ learn better user and item embeddings. 
Thus, we intend to verify the influence of two CL tasks on overall performance. 
Specifically, we conduct two different types of experiments: 
1) \textit{Parameter Sensitive Experiments}: investigating the impact of two CL tasks on the performance of \shortname;
2) \textit{Generalization of CL Tasks}: verifying the generalization of these two CL tasks on other review-based methods.
3) \textit{Quantitative Analysis about the Learned Interaction Features}: conducting quantitative analysis to evaluate the learned interaction features based on these two CL tasks.

\begin{table}
\caption{MSE comparison of \fusingshortname~ with different layers}
\begin{tabular}{@{}lccc@{}}
\toprule
\#Layers      & Digital\_Music                    & Toys\_and\_Games              & Clothing                   \\ \midrule
1 Layer     & 0.8037\small{$\pm$2e-3}           & 0.7853\small{$\pm$8e-4}       & 1.1024\small{$\pm$9e-4}    \\ 
2 Layers     & 0.8057\small{$\pm$1e-3}           & 0.7894\small{$\pm$2e-3}       & 1.1067\small{$\pm$3e-3}    \\ 
3 Layers     & 0.8123\small{$\pm$1e-3}           & 0.7942\small{$\pm$3e-3}       & 1.1144\small{$\pm$3e-3}    \\ 
\bottomrule
\end{tabular}
\label{tab:RGC_Layers}
\end{table}

\begin{figure}[t]

    \begin{subfigure}[t]{0.49\linewidth}
        \includegraphics[width=\textwidth]{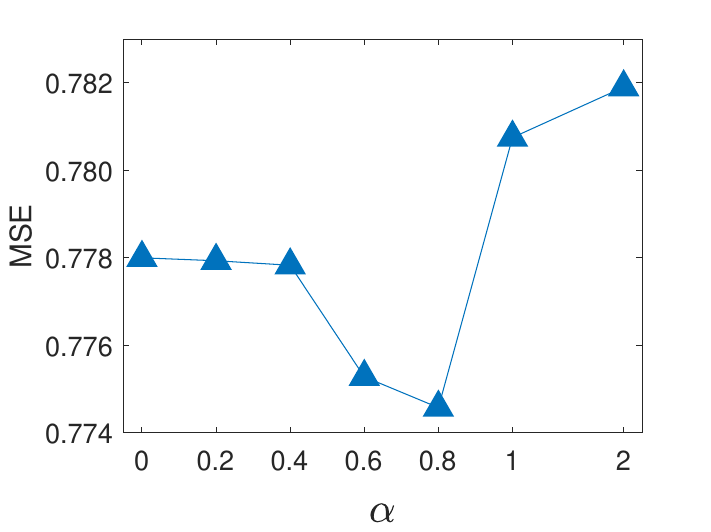}
        \caption{MSE w.r.t $\alpha$ on Digital Music}
    \end{subfigure}
    \begin{subfigure}[t]{0.49\linewidth}
        \includegraphics[width=\textwidth]{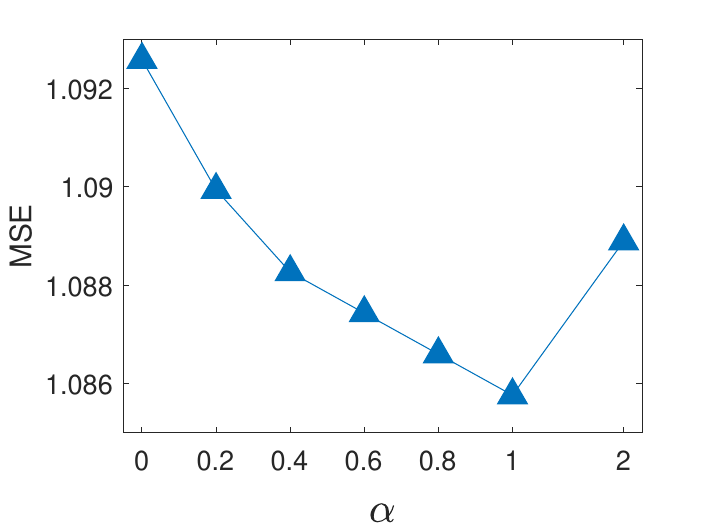}
        \caption{MSE w.r.t $\alpha$ on Clothing}
    \end{subfigure}
    
    \begin{subfigure}[t]{0.49\linewidth}
        \includegraphics[width=\textwidth]{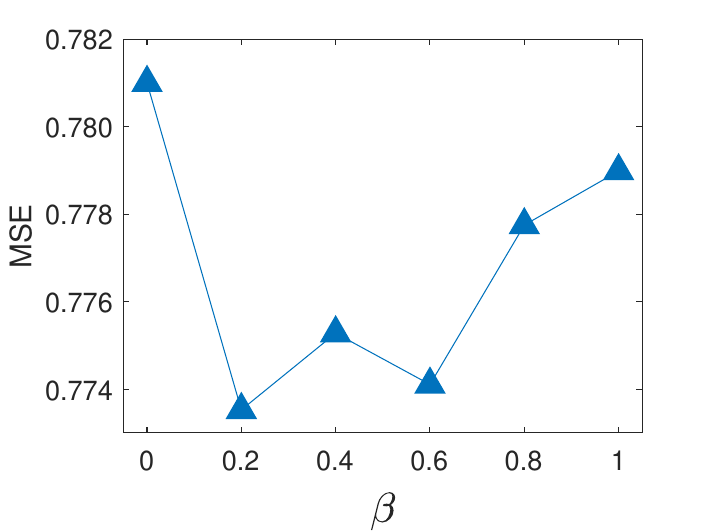}
        \caption{MSE w.r.t $\beta$ on Digital Music}
    \end{subfigure}
    \begin{subfigure}[t]{0.49\linewidth}
        \includegraphics[width=\textwidth]{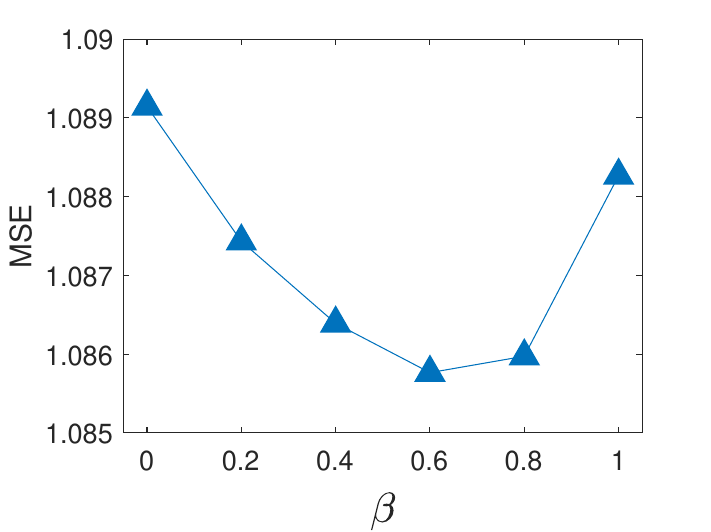}
        \caption{MSE w.r.t $\beta$ on Clothing}
    \end{subfigure}

    \caption{Rating prediction performance with different hyper-parameters on Digital Music and Clothing datasets.}
    \label{fig:Parameter_Analysis}
\end{figure}

\subsubsection{Parameter Sensitive Experiments}
To investigate the impact of ED and ND on the performance of \shortname, we conduct parameter sensitive experiments for two hyper-parameters $\alpha$ and $\beta$ on Digital\_Music and Clothing datasets.
The value of $\alpha$ is in the range  \{0, 0.2, 0.4, 0.6, 0.8, 1.0, 2.0\} and $\beta$ is in the range \{0, 0.2, 0.4, 0.6, 0.8, 1.0\}.
Corresponding results are summarized in \autoref{fig:Parameter_Analysis}. 

For the strength of ED loss (i.e., $\alpha$), we can obtain that model performance first increasing and then decreasing. The best values of $\alpha$ on these two datasets are $0.8$ and $1.0$, which is consistent with our assumption. ED can help the model focus on the consistent between user-item interaction and learned corresponding review embedding, which is very important for integrating review information for final prediction. 
Moreover, the target of \shortname~is to utilize learnt user and item embeddings to predict the final rating. 
If the value of $\alpha$ is too large, it will confuse the optimizing process of \shortname, which in turn limits the model performance. 

Moreover, we can obtain a similar conclusion of the strength of ND loss (i.e. $\beta$). The best values are $\beta=0.2$ for 'Digital Music' dataset and $\beta=0.6$ for 'Clothing' dataset. 
The choice of $\beta$ is distinct for different datasets, which depends on how easy the over-fitting happens caused by the specific data size and sparsity.

\subsubsection{Generalization of CL Tasks}
\label{s:generalization}
In order to better measure the generalization of CL tasks, we select review-based NARRE \cite{WWW2018NARRE} and graph-based GC-MC \cite{2017GC-MC} as backbones.
Then, we impose ED on NARRE to boost the interaction modeling. For GC-MC, we impose ND and ED tasks to verify the impact of both ND and ED tasks. 

In \autoref{tab:generalization}, ``+ED'' denotes imposing ED task into the model, and ``+ND+ED'' means incorporating two CL tasks into the model.
The observations are summarized as follows:
First of all, we can observe that both models benefit from our proposed ED and ND tasks, demonstrating the effectiveness and generalization of these two CL tasks. 
Second, the relative improvement on GC-MC is bigger than the improvement on NARRE across three datasets, demonstrating that ED helps accessing the necessary information during complex interaction modeling. 
Moreover, when incorporating both ED and ND into GC-MC, the performance of GC-MC can be further improved~($3.5\%, 2.0\%$, and $1.7\%$ improvement on three datasets).
This phenomenon verifies that ED and ND are essential for learning accurate user preferences.

\begin{table}[t]
\vspace{2mm}
\caption{MSE of our CL constraints on different recommendation models.}
\resizebox{1.\linewidth}{!}{
\begin{tabular}{@{}lccc@{}}
\toprule
Models      & Digital\_Music            & Toys\_and\_Games          & Clothing  \\ \midrule
NARRE       & 0.8172\small{$\pm$1e-3}    & 0.7962\small{$\pm$1e-3}   & 1.1064\small{$\pm$1e-3}  \\
\small{+ED}    & 0.8018\small{$\pm$4e-3} (1.9\%) & 0.7910\small{$\pm$1e-3}  (0.7\%)   & 1.0961\small{$\pm$1e-3}  (0.9\%)  \\ \midrule
GC-MC       & 0.8090\small{$\pm$1e-3}            & 0.7986\small{$\pm$5e-4}          & 1.1088\small{$\pm$1e-3}  \\
\small{+ED}     & 0.7822\small{$\pm$2e-3} (3.3\%)        & 0.7848\small{$\pm$2e-3} (1.7\%)  & 1.0922\small{$\pm$2e-3} (1.5\%)   \\
\small{+ND+ED}  & 0.7803\small{$\pm$1e-3} (3.5\%) & 0.7825\small{$\pm$1e-3}  (2.0\%)  & 1.0901\small{$\pm$6e-4}  (1.7\%)   \\ \bottomrule
\end{tabular}
}
\label{tab:generalization}
\end{table}

\begin{table}[t]
    \centering
    \caption{Comparisons of mutual information estimate between learned interaction representations $\{\bm{h}_{ij}|(i,j) \in \mathcal{E}\}$ and review representations $\{\bm{e}_{ij}|(i,j) \in \mathcal{E}\}$. The larger score means the smaller divergence.}
\begin{tabular}{@{}lccc@{}}
\toprule
Models    & Digital\_Music & Toys\_and\_Games & Clothing \\ \midrule
(1) TransNets & 0.09           & 0.06              & 0.03     \\
(2) SDNet     & 0.17           & 0.14              & 0.12     \\
(3) \fusingshortname       & 0.41           & 0.37              & 0.32     \\
(4) \fusingshortname+ND    & 1.35           & 1.22              & 0.86     \\
(5) \fusingshortname+ED    & 2.03           & 1.92              & 1.60     \\
(6) \shortname~      & 2.46           & 2.17              & 1.82     \\ \bottomrule 
\end{tabular}
\label{tab:MINE}
\end{table}

\subsubsection{Quantitative Analysis about the Learned Interaction Features.}
Previous experiments have proven the impact and generalization of CL tasks. 
In this section, we intend to make a quantitative analysis to measure the consistency between user-item interaction features and corresponding reviews. 
Since ED and ND are utilized to help \shortname~to learn better node representations, we focus on the quantitative evaluation of the learned representations. 
Specifically, following~\cite{DIM2018ICLR}, we also leverage MINE~\cite{MINE2018ICML} to estimate the KL-divergence between the learned user-item interaction $\bm{h}_{ij}$ and corresponding review representation $\bm{e}_{ij}$. 
We have compared the result of TransNets, SDNet, \fusingshortname+ED, and \fusingshortname+ND on three datasets, which have been reported in \autoref{tab:MINE}. Note that the bigger value indicates higher dependence. 

According to the results, we can observe the following phenomena: 
1) The interaction features generated from \shortname~have the best consistency performance, which is consistent with previous experimental results;
2) The comparison between \autoref{tab:MINE} (4) and (5) demonstrates that ED has a bigger impact on the model performance than ND, which is consistent with the results in Section~\ref{s:generalization}; 
3) Without CL tasks, our proposed \shortname~still performs better than TransNets and SDNet, indicating that our proposed \fusingshortname~is capable of extracting relevant information from reviews and graphs to model user preferences more accurately.

\section{Conclusion}
In order to fully exploit the unique structure of user-item bipartite graph with edge features from ratings and reviews, as well as employ review information to enhance user/item embedding and user-item interaction modeling, 
we proposed a novel \fullname~(\shortname), a graph-based contrastive learning framework for user preference modeling and rating prediction. 
Specifically, we designed a novel \fusingfullname~ module (\fusingshortname) to incorporate review information into user and item embedding learning more effectively. 
In this component, reviews were utilized to fine-tune the influences of corresponding neighbors and reviews themselves. 
Moreover, we developed two additional CL tasks~(i.e., ED and ND) to constrain \shortname~for better node embeddings and interaction modeling.
Finally, we have conducted extensive experiments over five benchmark datasets in recommendation to demonstrate the superiority and effectiveness of \shortname. 
In the future, we will incorporate more advanced review embedding methods for better review representation and design better fusing and alignment methods for complex interactions modeling between reviews and users (items). 

\begin{acks}
This work was supported in part by grants from the National Natural Science Foundation of China (Grant No. 72188101, 61725203, 91846201, 62006066), the Open Project Program of the National Laboratory of Pattern Recognition (NLPR), the Co-operative Innovation Project of Colleges in Anhui: GXXT-2019-025, CCF-AFSG Research Fund (Grant No. CCF-AFSG RF20210006), and the Young Elite Scientists Sponsorship Program by CAST and ISZ.
\end{acks}

\bibliographystyle{ACM-Reference-Format}
\bibliography{reference}

\end{document}